 \documentclass[final,5p,times,twocolumn]{elsarticle}

\usepackage{amssymb}

\usepackage{upgreek}
\usepackage{color}
\usepackage{amsmath}
\usepackage{lineno}
\biboptions{numbers,sort&compress}
\newcommand{\eplus}{\ensuremath{\mathrm{e}^{+}}}
\newcommand{\eminus}{\ensuremath{\mathrm{e}^{-}}}

\newcommand{\photon}{\ensuremath{\upgamma}}

\newcommand{\muplus}{\ensuremath{\upmu^+}}
\newcommand{\muminus}{\ensuremath{\upmu^-}}

\newcommand{\piplus}{\ensuremath{\uppi^+}}
\newcommand{\piminus}{\ensuremath{\uppi^-}}
\newcommand{\pizero}{\ensuremath{\uppi^0}}
\newcommand{\uboson}{\ensuremath{\mathrm{U}}}

\newcommand{\Aprime}{\ensuremath{{\mathrm{A}^{\prime}}}}
\newcommand{\gprime}{\ensuremath{{\mathrm{\upgamma}^{\prime}}}}

\newcommand{\rhomeson}{\ensuremath{\uprho}}
\newcommand{\wmeson}{\ensuremath{\upomega}}

\newcommand{\eeUgUpp}{\ensuremath{\eplus\eminus \to \uboson\photon\mathrm{,}\,\, \uboson \to \piplus\piminus}}

\newcommand{\eV}{{e\kern-.06em V}}

\newcommand{\MeVc}{{\rm \,M\eV\kern-.05em /\kern-.02em c}}
\newcommand{\MeVcc}{{\rm \,M\eV\kern-.05em /\kern-.02em c$^{2}$}}

\newcommand{\GeVcc}{{\rm \,G\eV\kern-.05em /\kern-.02em c$^{2}$}}
\def\ifm#1{\relax\ifmmode#1\else$#1$\fi}  \def\to{\ifm{\rightarrow}}  \def\epm{\ifm{e^+e^-}}
\def\gam{\ifm{\gamma}}     \def\x{\ifm{\times}}  \def\ab{\ifm{\sim}}    
\def\up#1;{$^{#1}$}  \def\dn#1;{$_{#1}$}    \def\DAF{DA\char8NE}  
  \def\pic{\ifm{\pi^+\pi^-}}  \def\po{\ifm{\pi^0}}
\def\pt#1;#2;{\ifm{#1\x10^{#2}}}  
\def\mt{\ifm{M_{\rm trk}}}

\newcommand{\affuni}[2]{Dipartimento di Fisica dell'Universit\`a #1, #2, Italy.}
\newcommand{\affinfn}[2]{INFN Sezione di #1, #2, Italy.}

\journal{Physics Letters B}

\begin{document}
\begin{frontmatter}



\title{Limit on the production of a new vector boson in $\eeUgUpp $ with the KLOE experiment}

\author{The KLOE-2 Collaboration}
\author[Messina,Frascati]{\\ A.~Anastasi}
\author[Frascati]{D.~Babusci}
\author[Frascati]{G.~Bencivenni}
\author[Warsaw]{M.~Berlowski}
\author[Frascati]{C.~Bloise}
\author[Frascati]{F.~Bossi}
\author[INFNRoma3]{P.~Branchini}
\author[Roma3,INFNRoma3]{A.~Budano}
\author[Uppsala]{L.~Caldeira~Balkest\aa hl}
\author[Uppsala]{B.~Cao}
\author[Roma3,INFNRoma3]{F.~Ceradini}
\author[Frascati]{P.~Ciambrone}
\author[Messina,INFNCatania,Novosibirsk]{F.~Curciarello\corref{mycorrespondingauthor}}
\cortext[mycorrespondingauthor]{Corresponding author}
\ead{fcurciarello@unime.it}
\author[Cracow]{E.~Czerwi\'nski}
\author[Roma1,INFNRoma1]{G.~D'Agostini}
\author[Frascati]{E.~Dan\`e}
\author[INFNRoma3]{V.~De~Leo}
\author[Frascati]{E.~De~Lucia}
\author[Frascati]{A.~De~Santis}
\author[Frascati]{P.~De~Simone}
\author[Roma3,INFNRoma3]{A.~Di~Cicco}
\author[Roma1,INFNRoma1]{A.~Di~Domenico}
\author[INFNRoma2]{R.~Di~Salvo}
\author[Frascati]{D.~Domenici}
\author[Frascati]{A.~D'Uffizi}
\author[Roma2,INFNRoma2]{A.~Fantini}
\author[Frascati]{G.~Felici}
\author[ENEACasaccia,INFNRoma1]{S.~Fiore}
\author[Cracow]{A.~Gajos}
\author[Roma1,INFNRoma1]{P.~Gauzzi}
\author[Messina,INFNCatania]{G.~Giardina}
\author[Frascati]{S.~Giovannella}
\author[INFNRoma3]{E.~Graziani}
\author[Frascati]{F.~Happacher}
\author[Uppsala]{L.~Heijkenskj\"old}
\author[Uppsala]{W.~Ikegami Andersson}
\author[Uppsala]{T.~Johansson}
\author[Cracow]{D.~Kami\'nska}
\author[Warsaw]{W.~Krzemien}
\author[Uppsala]{A.~Kupsc}
\author[Roma3,INFNRoma3]{S.~Loffredo}
\author[Messina2,INFNMessina]{G.~Mandaglio \corref{mycorrespondingauthor}}
\ead{gmandaglio@unime.it}
\author[Frascati,Marconi]{M.~Martini}
\author[Frascati]{M.~Mascolo}
\author[Roma2,INFNRoma2]{R.~Messi}
\author[Frascati]{S.~Miscetti}
\author[Frascati]{G.~Morello}
\author[INFNRoma2]{D.~Moricciani}
\author[Cracow]{P.~Moskal}
\author[Boston]{A.~Palladino}
\author[Uppsala]{M.~Papenbrock}
\author[INFNRoma3]{A.~Passeri}
\author[Energetica,INFNRoma1]{V.~Patera}
\author[Frascati]{E.~Perez~del~Rio}
\author[INFNBari]{A.~Ranieri}
\author[Frascati]{P.~Santangelo}
\author[Frascati]{I.~Sarra}
\author[Calabria,INFNCalabria]{M.~Schioppa}
\author[Frascati]{M.~Silarski}
\author[Frascati]{F.~Sirghi}
\author[INFNRoma3]{L.~Tortora}
\author[Frascati]{G.~Venanzoni}
\author[Warsaw]{W.~Wi\'slicki}
\author[Uppsala]{M.~Wolke}
\address[INFNBari]{\affinfn{Bari}{Bari}}
\address[INFNCatania]{\affinfn{Catania}{Catania}}
\address[Cracow]{Institute of Physics, Jagiellonian University, Cracow, Poland.}
\address[Frascati]{Laboratori Nazionali di Frascati dell'INFN, Frascati, Italy.}
\address[Messina]{Dipartimento di Scienze Matematiche e Informatiche, Scienze Fisiche e Scienze della Terra dell'Universit\`a di Messina, Messina, Italy.}
\address[Messina2]{Dipartimento di Scienze Chimiche, Biologiche, Farmaceutiche ed Ambientali dell'Universit\`a di Messina, Messina, Italy.}
\address[INFNMessina]{INFN Gruppo collegato di Messina, Messina, Italy.}
\address[Calabria]{\affuni{della Calabria}{Rende}}
\address[INFNCalabria]{INFN Gruppo collegato di Cosenza, Rende, Italy.}
%
\address[Energetica]{Dipartimento di Scienze di Base ed Applicate per l'Ingegneria dell'Universit\`a 
``Sapienza'', Roma, Italy.}
\address[Marconi]{Dipartimento di Scienze e Tecnologie applicate, Universit\`a ``Guglielmo Marconi", Roma, Italy.}
\address[Novosibirsk]{Novosibirsk State University, 630090 Novosibirsk, Russia.}
\address[Roma1]{\affuni{``Sapienza''}{Roma}}
\address[INFNRoma1]{\affinfn{Roma}{Roma}}
\address[Roma2]{\affuni{``Tor Vergata''}{Roma}}
\address[INFNRoma2]{\affinfn{Roma Tor Vergata}{Roma}}
\address[Roma3]{Dipartimento di Matematica e Fisica dell'Universit\`a 
``Roma Tre'', Roma, Italy.}
\address[INFNRoma3]{\affinfn{Roma Tre}{Roma}}
\address[ENEACasaccia]{ENEA UTTMAT-IRR, Casaccia R.C., Roma, Italy}
\address[Boston]{Department of Physics, Boston University, Boston, USA}
\address[Uppsala]{Department of Physics and Astronomy, Uppsala University, Uppsala, Sweden.}
\address[Warsaw]{National Centre for Nuclear Research, Warsaw, Poland.}

\begin{abstract}
The recent interest in a light gauge boson in the framework of an extra U(1) symmetry motivates searches in the mass range below 1 GeV.
 We present a search for such a particle, the dark photon, in $\eeUgUpp$ based on 28 million $\mathrm{e^+ e^-} \to \uppi^+ \uppi^-\upgamma$ events collected at \DAF\ by the KLOE experiment. The \piplus\ \piminus\ production by initial-state radiation compensates for a loss of sensitivity of previous KLOE \uboson\ \to\ \eplus\ \eminus\ , \muplus\ \muminus\ searches due to the small branching ratios in the $\rhomeson-\wmeson$  resonance region. 
We found no evidence for a signal and set a limit at 90\% CL on the mixing strength between the photon and the dark photon, $\varepsilon^2$, in the U mass range between $527$ and $987$~MeV.  Above 700 MeV this new limit is more stringent than previous ones.
\end{abstract}


\end{frontmatter}


\section{Introduction}
\label{sec:intro}
A new kind of matter, called dark matter (DM), which does not absorb or emit light, has been postulated since the early '30s of the past century~\cite{Zwicky} and its existence is now widely accepted~\cite{PDG}. However, its interpretation is still among the greatest and fascinating enigmas of Physics.
The current paradigm assumes that the DM is a thermal relic from the Big Bang, accounting for about 24\% of the total energy density of the Universe~\cite{PDG} and producing effects only through its gravitational interactions with large-scale cosmic structures. 
To include the DM in a particle theoretical framework, the Standard Model (SM) is usually complemented with many extensions~\cite{Holdom,U_th1, Fayet,U_th2,U_th6} that attribute to the DM candidates also strong self interactions and weak-scale interactions with SM particles. Among the possible candidates,  a Weakly Interacting Massive Particle (WIMP) aroused much interest since a particle with weak-scale annihilation cross section can account for the DM relic abundance estimated through the study of the cosmic microwave background~\cite{PDG}.
The force carrier in WIMP annihilations could be a new gauge vector boson, known  as \uboson\ boson, dark photon, $\gprime$ or $\Aprime$, with allowed decays into leptons and hadrons. Its assiduous worldwide search has been  strongly motivated by the astrophysical evidence recently observed in many experiments~\cite{Pamela,AMS,Integral,Atic,Hess,Fermi,Dama/Libra} and by its possible positive one-loop contribution to the theoretical value of the muon magnetic moment anomaly~\cite{a_mu}, which could solve, partly or entirely, the well known 3.6~$\sigma$ discrepancy with the experimental measurement~\cite{a_mu1}.

In this paper we assume the simplest theoretical hypothesis according to which the dark sector consists of just one extra abelian gauge symmetry, $\rm U(1)$, with one gauge boson, the U boson, whose decays into invisible light dark matter are kinematically inaccessible. In this framework the dark photon  would act like a virtual photon, with virtuality $q^2 = m_{\uboson}^2$. It would couple to leptons and quarks with the same strength  and would appear (its width being much smaller than the experimental mass resolution) as a narrow resonance in any process involving real or virtual photons.
   The coupling to the SM photon would occur by means of a vector portal mechanism~\cite{Holdom}, i.e. loops of heavy dark particles charged under both the SM and the dark force. The strength of the mixing with the photon is parametrized  by a single factor $\varepsilon^2 = \alpha^{'}/\alpha$ which is the ratio of the effective dark and SM photon couplings~\cite{Holdom}. The size of the $\varepsilon^2$ parameter is expected to be very small  ($10^{-2}-10^{-8}$) causing a suppression of the U boson production rate. The U boson decays into SM particles would happen through the same mixing operator, with the corresponding decay amplitude suppressed by an $\varepsilon^2$ factor, but are still expected to be detectable at high luminosity $\eplus \eminus$ colliders~\cite{Essig,Batell,Reece}.
   
KLOE has already searched for radiative U boson production in the $\mathrm{e}^+ \mathrm{e}^- \to \rm \uboson \upgamma\,, \uboson\to \mathrm{e}^+ \mathrm{e}^-,\,  \upmu^+  \upmu^-$ processes~\cite{eeg,mmg}. 
The leptonic channels are affected by a decrease in sensitivity in the $\uprho-\upomega$ region due to the dominant branching fraction into hadrons.
For a virtual photon with $q^2 < 1$~GeV$^{2}$, the coupling to  the charged pion is given by the product of the electric charge and the pion form factor, $e\, F_{\pi}(q^2)$. The effective coupling of the U boson to pions is thus predicted to be given by the product of the virtual photon coupling and the kinetic mixing parameter $\varepsilon\, e\, F_{\pi}(q^2)$~\cite{Reece}.
Being far from the $\uppi^+ \uppi^-$ mass threshold (see Section~\ref{sec:selection}) finite mass effects can be safely neglected.
We thus searched for a short lived U boson decaying to $\uppi^+\uppi^-$ in a data sample corresponding to 1.93~fb$^{-1}$ integrated luminosity, by looking for a resonant peak in the dipion invariant mass spectrum with initial-state radiation (ISR) $\uppi^+\uppi^-\upgamma$ events.

\section{ The KLOE detector}
\label{sec:kloe_detector}

The KLOE detector operates at \DAF\,, the Frascati $\upphi$-factory. \DAF\ is an \eplus \eminus\  collider usually operated at a center of mass energy $ m_\upphi\simeq 1.019$ GeV. Positron and electron beams collide at an angle of $\pi-$25 mrad, producing $\upphi$ mesons nearly at rest. The KLOE detector consists of a large cylindrical drift chamber (DC)~\cite{KLOE_DC}, surrounded by a lead scintillating-fiber electromagnetic calorimeter (EMC)~\cite{KLOE_EMC}.  
A superconducting coil around  the EMC provides a 0.52 T magnetic field along the bisector of the colliding beams. The bisector is taken as the $z$ axis of our coordinate
system. The $x$ axis is horizontal, pointing to the center of the
collider rings and the $y$ axis is vertical, directed upwards.

The EMC barrel and end-caps cover 98\% of the solid angle. Calorimeter modules are read out at both ends by 4880 photomultipliers. Energy and time resolutions are $ \sigma_E /E=~0.057 /\sqrt{E(\rm{GeV})} $ and $ \sigma_t =57\ \rm{ps}/\sqrt{E(\rm{GeV})}\oplus 100\ \rm{ps}$, respectively. 
The drift chamber has only stereo wires and is $4$ m in diameter, $ 3.3$ m long. It is built out of carbon-fibers and operates with a low-$Z$ gas mixture (helium with 10\% isobutane). Spatial resolutions are  $\sigma_{xy}\ab150\ \rm\upmu m$ and  $\sigma_z\ab2$ mm. The momentum resolution for large angle tracks is $\sigma(p_\perp) / p_\perp\ab 0.4\% $. The trigger uses both EMC and DC information. Events used in this analysis are triggered by at least two energy deposits larger than 50~MeV in two sectors of the barrel calorimeter~\cite{KLOE_trig}.

\section{Event selection}
\label{sec:selection}

We selected $\piplus \piminus \upgamma$ candidates by requesting events with two oppositely-charged tracks emitted at large polar angles, $50^{\circ}~<~\theta~<~130^{\circ}$, with the  undetected ISR photon missing momentum pointing -- according to the $\piplus \piminus \upgamma$ kinematics --  at a small polar angles ($\theta~<~15^{\circ},\, \theta~>~165^{\circ}$).
The tracks were required to have the point of closest approach to the $z$  axis within a cylinder of radius 8 cm and length 15 cm centred at the interaction point. 
In order to ensure good reconstruction and efficiency, we selected tracks with transverse and longitudinal momentum in the range $\mathrm{p}_{\perp}>$~160~MeV or $\mathrm{p}_{\parallel} > $~90~MeV, respectively.

Since the $\uppi^+\uppi^- \upgamma$ cross section behaviour as a function of the ISR photon polar angle is divergent ($\propto 1/\theta_{\upgamma}^4$), the track and the photon acceptance selections make the final-state radiation (FSR) and the $\upphi$ resonant processes relatively unimportant, leaving us with a high purity ISR sample and  increasing our sensitivity to the $\rm U \to \uppi^+ \uppi^- $ decay~\cite{Babayaga}.

\begin{figure}[htb]
\begin{center}
\includegraphics[width=8cm]{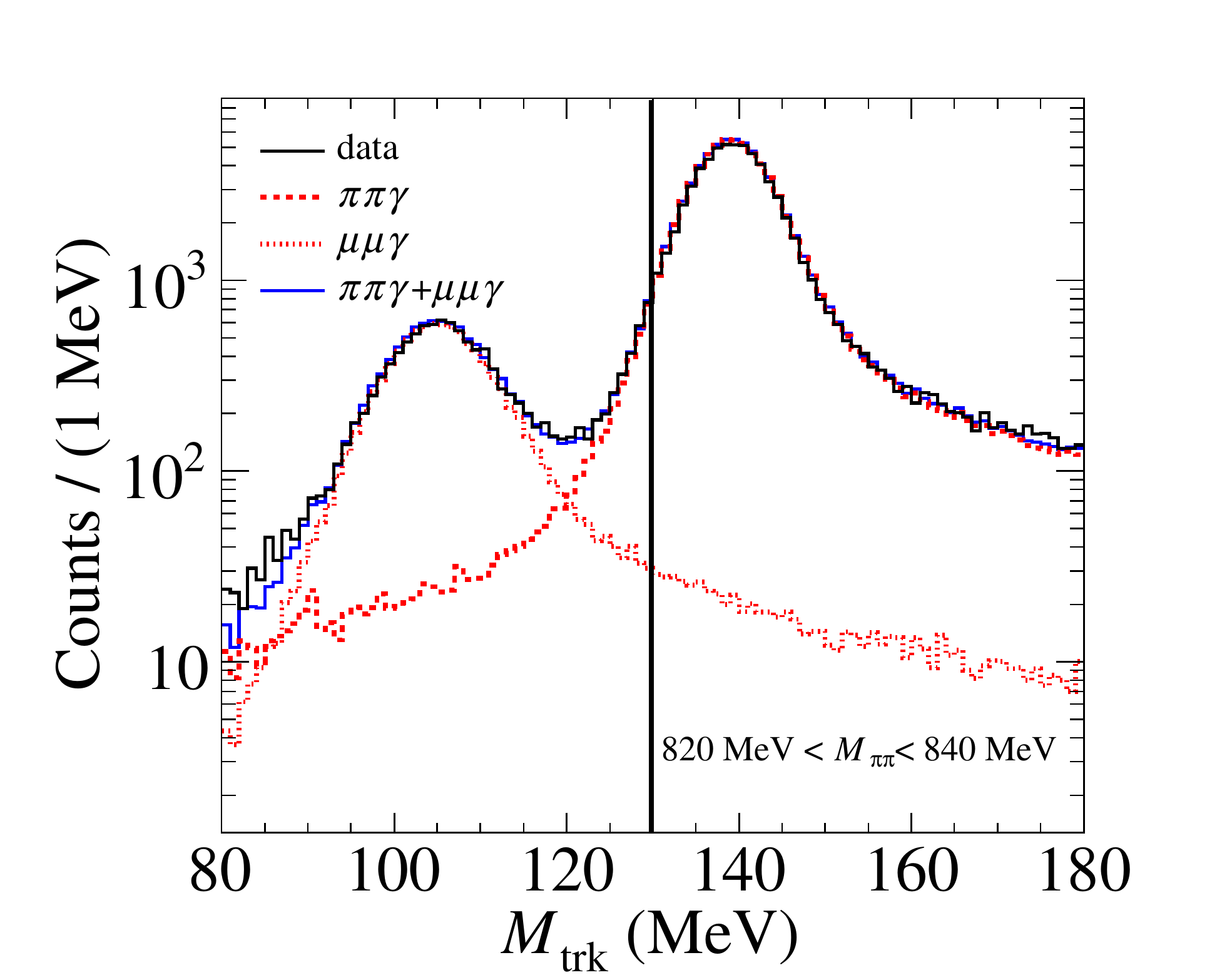}
\caption{Example of $M_{\rm trk}$ distributions for the $M_{\uppi\uppi}=820-840$ MeV bin. Measured data are represented in black, simulated $ \uppi^+ \uppi^-\upgamma$ and $ \upmu^+ \upmu^-\upgamma$ in red. Simulated $ \upmu^+ \upmu^-\upgamma+\uppi^+\uppi^-\upgamma$ in blue. Events at the left of the vertical line are rejected.}
\label{mtrk}
\end{center}
\end{figure}

The Monte Carlo simulation of the $M_{\pi\pi}$ spectrum was produced with the  $\mathtt{PHOKHARA}$ event generator ~\cite{PHOKHARA} with the K\"{u}hn--Santamaria (KS)~\cite{kusa} pion form factor parametrization and included a full description of the KLOE detector (GEANFI package~\cite{GEANFI}). The collected data were simulated including $\upphi$ decays and leptonic processes $\eplus\eminus \to  \ell^+ \ell^- \upgamma(\upgamma),\, \ell=\mathrm{e},\upmu$.
The main background contributions affecting the ISR $\uppi^+\uppi^-\upgamma$ sample are the resonant $ \mathrm{e}^+ \mathrm{e}^- \to\upphi \to  \uppi^+  \uppi^-  \uppi^0$ process and the ISR and FSR $\eplus\eminus \to  \ell^+ \ell^- \upgamma(\upgamma),\, \ell=\mathrm{e},\upmu$ processes (they will be defined as ``residual background" in the following). 
We reduced their contribution by applying kinematic cuts in the $M_{\mathrm trk}-M^2_{ \uppi \uppi}$ plane, as explained in Refs.~\cite{KLOE1,KLOE2}. $M^2_{ \uppi \uppi}$ is the squared invariant mass of the two selected tracks in the pion mass hypothesis while \mt\  is the mass of the charged particles associated to the tracks, computed in the equal mass hypothesis and assuming that the missing momentum of the event pertains to a single photon. 
Distributions of the $M_{\mathrm trk}$ variable for data and simulation are shown in Fig.~\ref{mtrk}, where the $M_{\mathrm trk}> 130$~MeV cut to discriminate muons from pions is also indicated.

A particle ID estimator (PID), $L_{\pm}$, defined for each track with associated energy released in EMC and based on a pseudo-likelihood function, uses calorimeter information (size and shape of the energy depositions and time of flight) to suppress radiative Bhabha scattering events ~\cite{memo,KLOE1,KLOE2}.
\begin{figure}[htb!]
\begin{center}\vspace{-0.2cm}
\includegraphics[width=8cm]{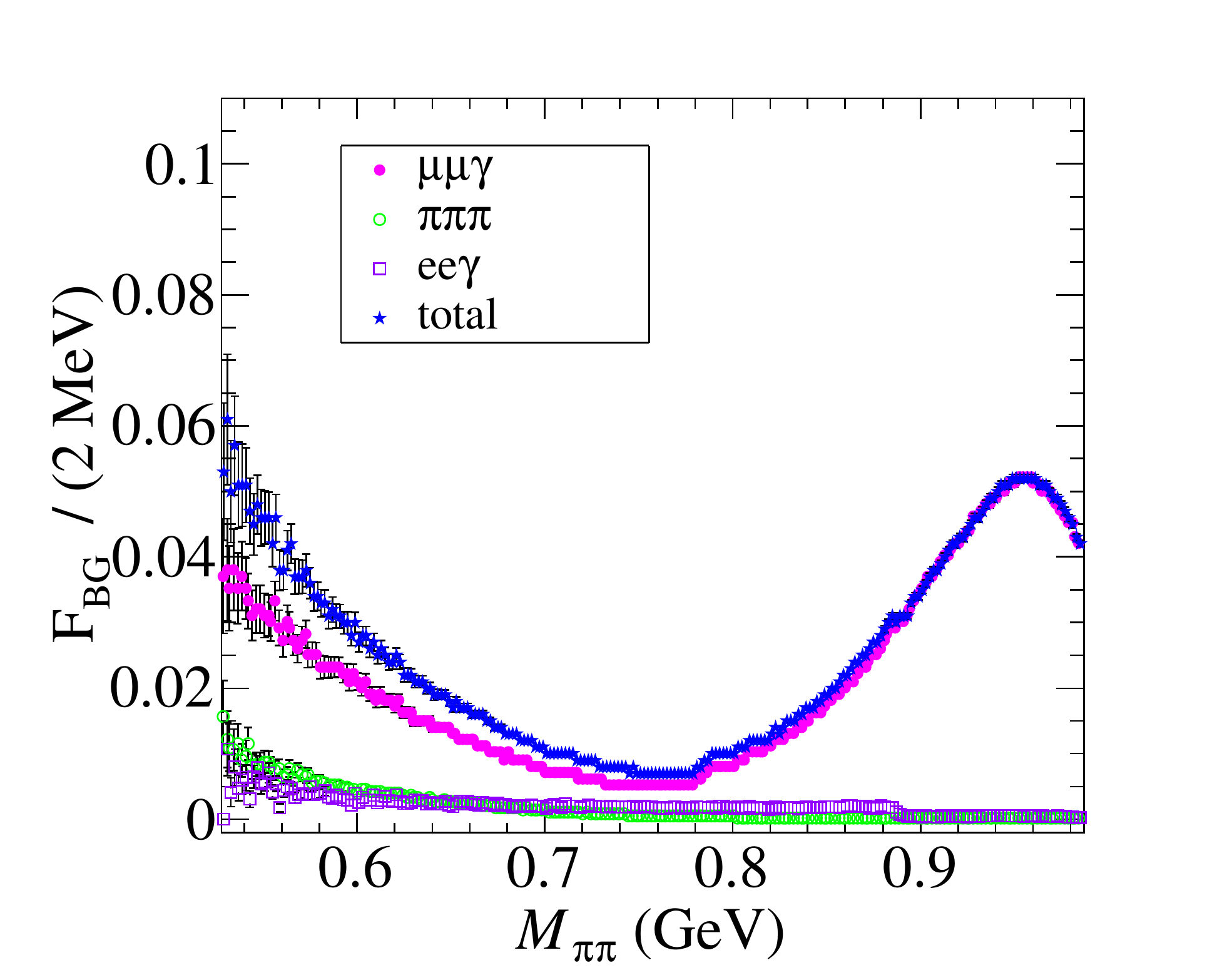}
\caption{Fractional backgrounds, normalized to the $\uppi^+\uppi^-\upgamma$ contribution, from the $\pic\po$, $\epm\gam$, and $\muplus \muminus \upgamma$ channels after all selection criteria.}
\label{Bckg}
\end{center}
\end{figure}

Electrons deposit their energy mainly at the entrance of the calorimeter  while muons and pions tend to have a deeper penetration in the EMC. Events with both tracks having $L_{\pm} < 0$ are identified as $\mathrm{e}^+  \mathrm{e}^- \upgamma$ events and rejected. The efficiency of this selection is larger than 99.95\% as evaluated using measured data and simulated $\uppi^+\uppi^-\upgamma$ samples.

After these selections, about $2.8\times 10^7$ events are left in the measured data sample. We then applied the same analysis chain to the Monte Carlo simulated data: most of the selected sample consists of $\uppi^+\uppi^-\upgamma$ events, with residual ISR $\ell^+ \ell^- \upgamma,\, \ell=\mathrm{e},\upmu$ and $\upphi \to \uppi^+ \uppi^- \uppi^0$. Figure~\ref{Bckg} shows the fractional components of the residual background, $F_{\rm BG}$, individually for each contributing channel and their sum. The residual background rises up to about 6\% at low invariant masses and to 5\% above 0.9 GeV, decreasing to less than 1\% in the resonance region. 
\begin{figure}[h!]
\begin{center}
\includegraphics[width=8cm]{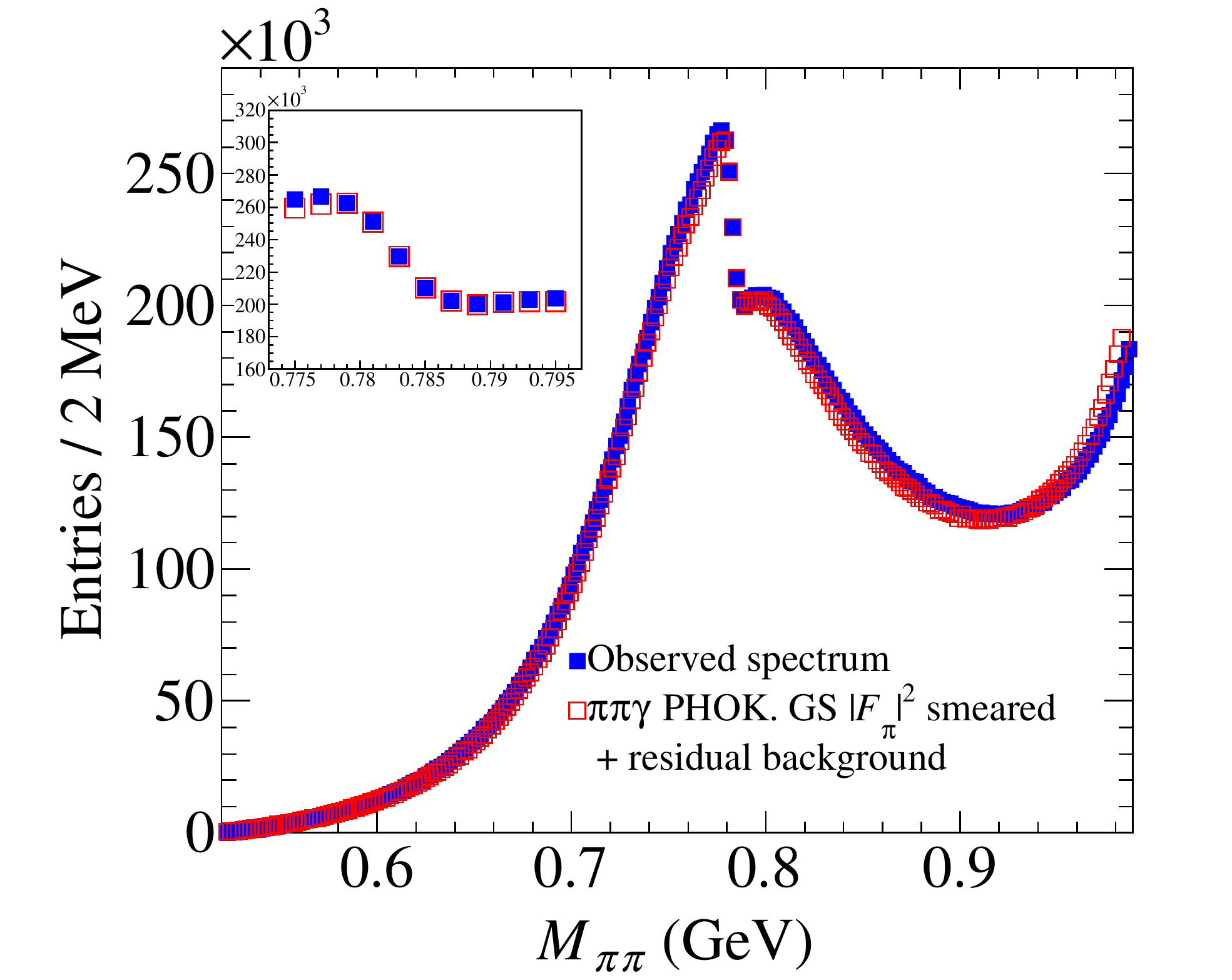}
\caption{Comparison of measured data (blue squares) and simulation performed  with the Gounaris-Sakurai $|F_{\pi}|^2$ parametrization (red open squares) for the $M_{\pi\pi}$ invariant mass spectrum. 
The figure insert shows in detail the agreement achieved in the $\rho-\omega$ mixing region (779--791 MeV). 
}
\label{data_mc}
\end{center}
\end{figure}

 A  very good description of the $\rho-\omega$ interference region (see the insert of Fig. 3) was achieved  by producing a dedicated sample using $\mathtt{PHOKHARA}$ as event generator with the Gounaris-Sakhurai (GS) pion form factor parametrization~\cite{Gosa}. The generation process used properly smeared distributions in order to account for the dipion invariant mass resolution (1.4--1.8~MeV).
In  Fig.~\ref{data_mc} the measured data spectrum is compared with the results of this simulation process, which includes the residual background.

\section{Irreducible background parametrization and estimate}
\label{sec:bkgrd}

Except for the $\rho-\omega$ region, we estimated the irreducible background directly from data.
For each U mass hypothesis the data are fitted in a $M_{\pi\pi}$ interval centred at $M_{\rm U}$ and 18-20 times wider than the $M_{\rm \pi \pi}$ resolution $\sigma_{M_{\rm \pi \pi}}$. The background is modelled by a monotonic function using Chebyshev polinomials up to the sixth order and is estimated using the sideband technique, by excluding from the fit the data in the region $\pm 3\, \sigma_{M_{\rm \pi \pi}}$ around $M_{\rm U}$~\cite{eeg}. The procedure is repeated in steps of 2 MeV in $M_{\rm U}$.

\begin{figure}[h]
\begin{center}
\includegraphics[width=8cm]{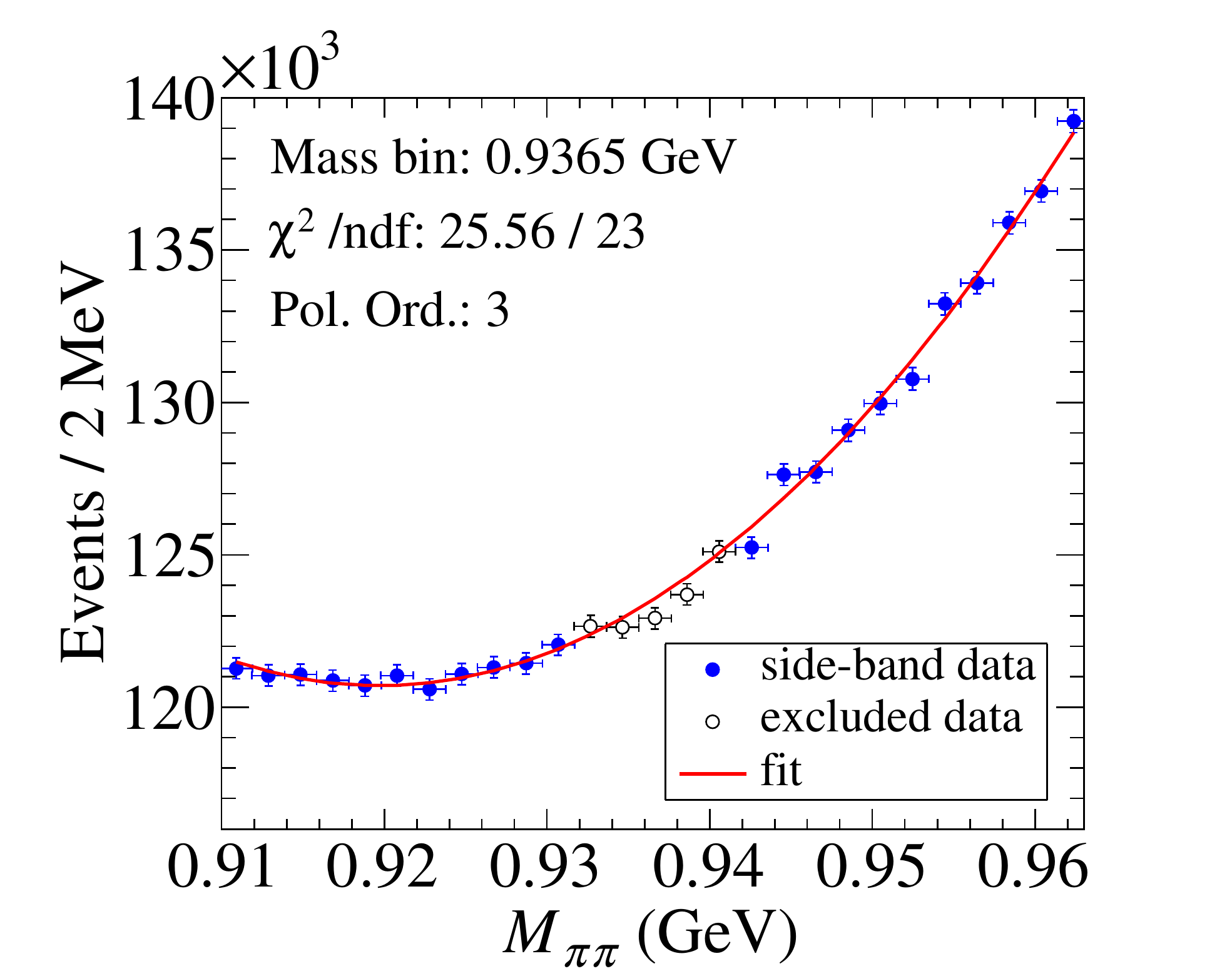}
\caption{Example of a Chebyshev polinomial sideband fit for the $M_{\rm U}=936$~MeV hypothesis}.
\label{sbfit}
\end{center}
\end{figure}


\begin{figure}[htp!]
\begin{center}
\includegraphics[width=8cm]{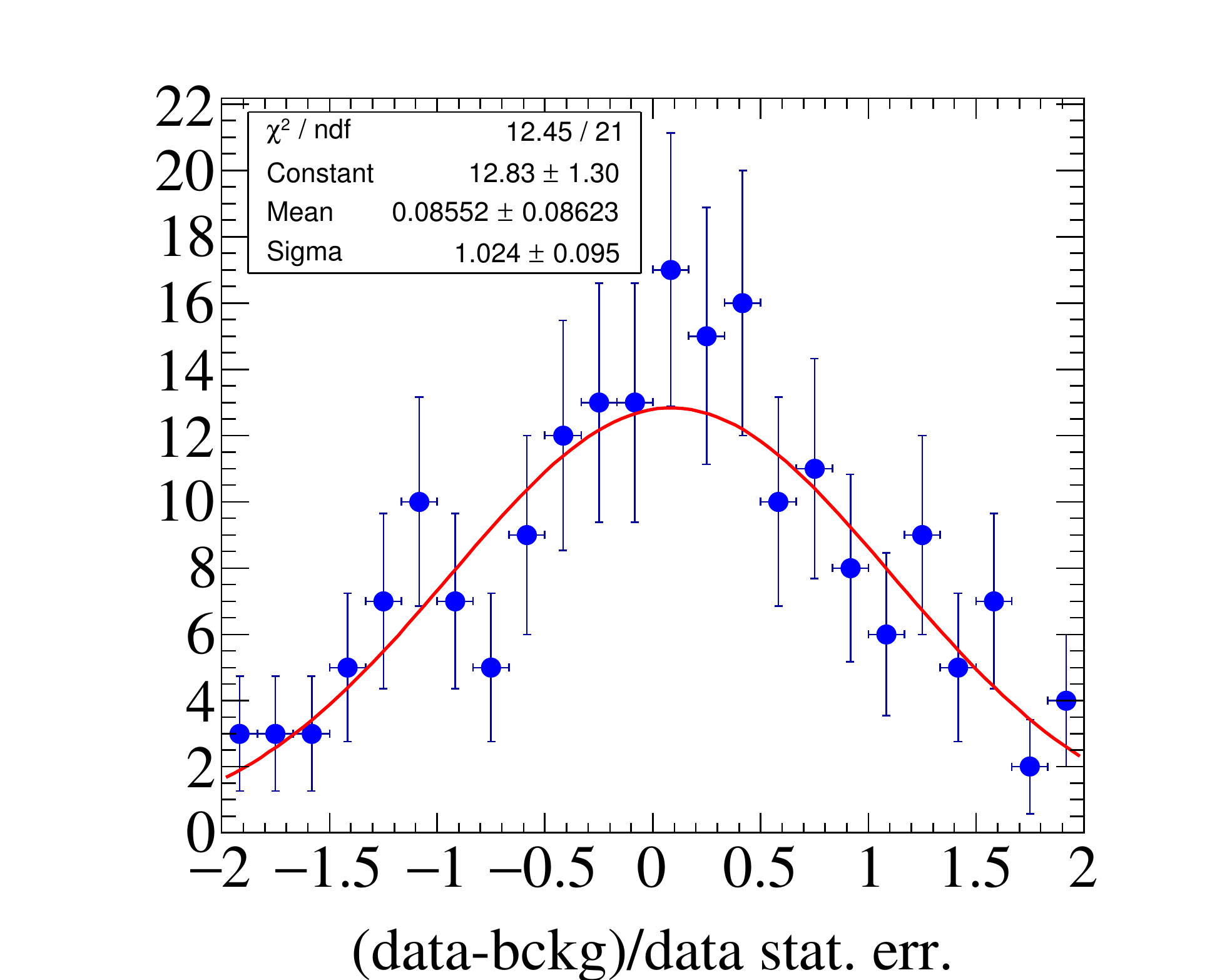}
\caption{Distribution of the differences (pulls) data-background normalized to the data statistical error (blue points) and relative Gaussian fit (red curve).}
\label{gauss}
\end{center}
\end{figure}
Fits with the best reduced $\chi^2$ are selected as histograms representing the background.  
For all used mass intervals, the distributions were  found to be smooth, with no ``wiggles" in any mass sub-range.
An example of the fit procedure is reported in Fig. \ref{sbfit}.
Figure~\ref{gauss} shows the distribution of the differences (pulls) between data and the  fitted background normalized to the data statistical error. Also shown is a Gaussian fit of this distribution.
 The mean and width parameters of the Gaussian fit are around zero and one, respectively.

The region of  $\rho-\omega$ interference is not smooth (see Fig.~\ref{data_mc}) and then not easy to be fitted with the sideband technique. We thus estimated the background in this region by using the  $\mathtt{PHOKHARA}$ generator with smeared distributions, as explained in Section~\ref{sec:selection} and shown in Fig.~\ref{data_mc} for the 779--791 MeV mass range.

\section{Systematic errors and efficiencies}
\label{sec:syst}

The main systematic uncertainties affecting this analysis are related to the evaluation of the irreducible background. As two different procedures were used in different mass ranges, the estimate of the systematic error accounted for two independent sources:
\begin{itemize}  
\item systematic uncertainties due to the sideband fitting procedure;
\item systematic uncertainties due to the evaluation of the background with  the $\mathtt{PHOKHARA}$ generator and to the smearing procedure in the 779-791 MeV mass range.
\end{itemize}

\begin{figure}[htb!]
\begin{center}
\includegraphics[width=8cm]{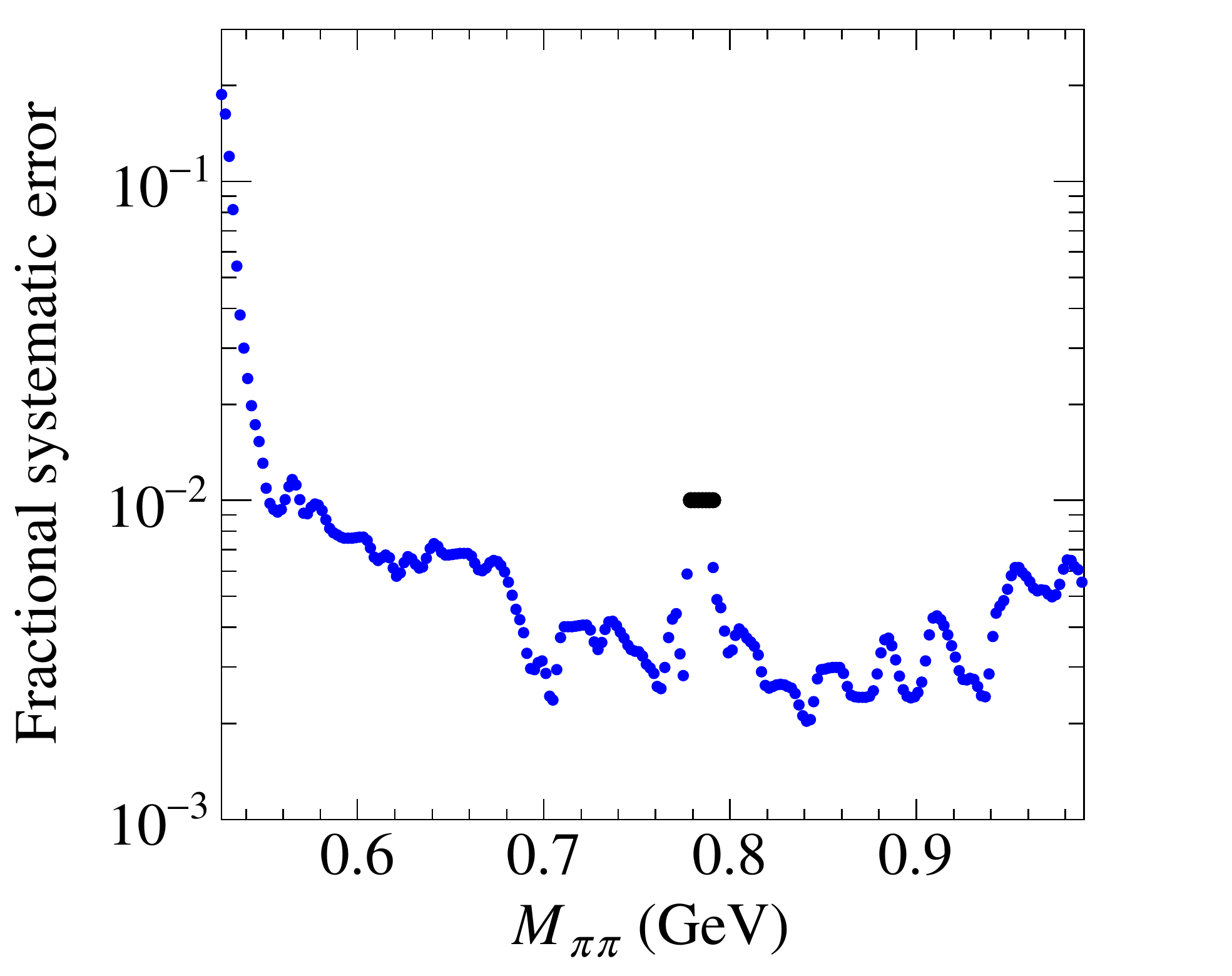}

\caption{ Fractional systematic error on the estimated background. Blue points: errors from the sideband fit procedure; black points: errors estimated from the $\mathtt{PHOKHARA}$ Monte Carlo simulation  for the $\rho-\omega$ interference region.}
\label{tot_tlimit_syst}
\end{center}
\end{figure}
The evaluation of the systematic uncertainties on the fitted background was performed by adding in quadrature, bin by bin, the contributions due to the errors of the fit and a systematic error due to the fit procedure. The first is obtained by propagating, for each fit interval, the corresponding errors of the fit parameters. The second is evaluated by varying the fit parameters by $\pm1\sigma$ and computing the maximum difference between the standard fit and the fit derived by using the modified parameters. The systematic error is less than 1\% in most of the mass range. 

In the $\rho-\omega$ region the systematic error is computed by adding in quadrature the contributions due to the theoretical uncertainty of the Monte Carlo generator ($0.5\%$~\cite{PHOKHARA}), the systematic error due to the residual background evaluation (0.3~\%, computed by changing the analysis cuts within the corresponding experimental resolutions), the contribution of the smearing procedure (0.8\%, obtained by varying the applied smearing of $\pm1\,\sigma$), and the systematic uncertainty on the luminosity (0.3~\%~\cite{PHOKHARA}). The resulting total systematic error is about 1\%.

\begin{figure}[htp!]
\begin{center}
\includegraphics[width=8cm]{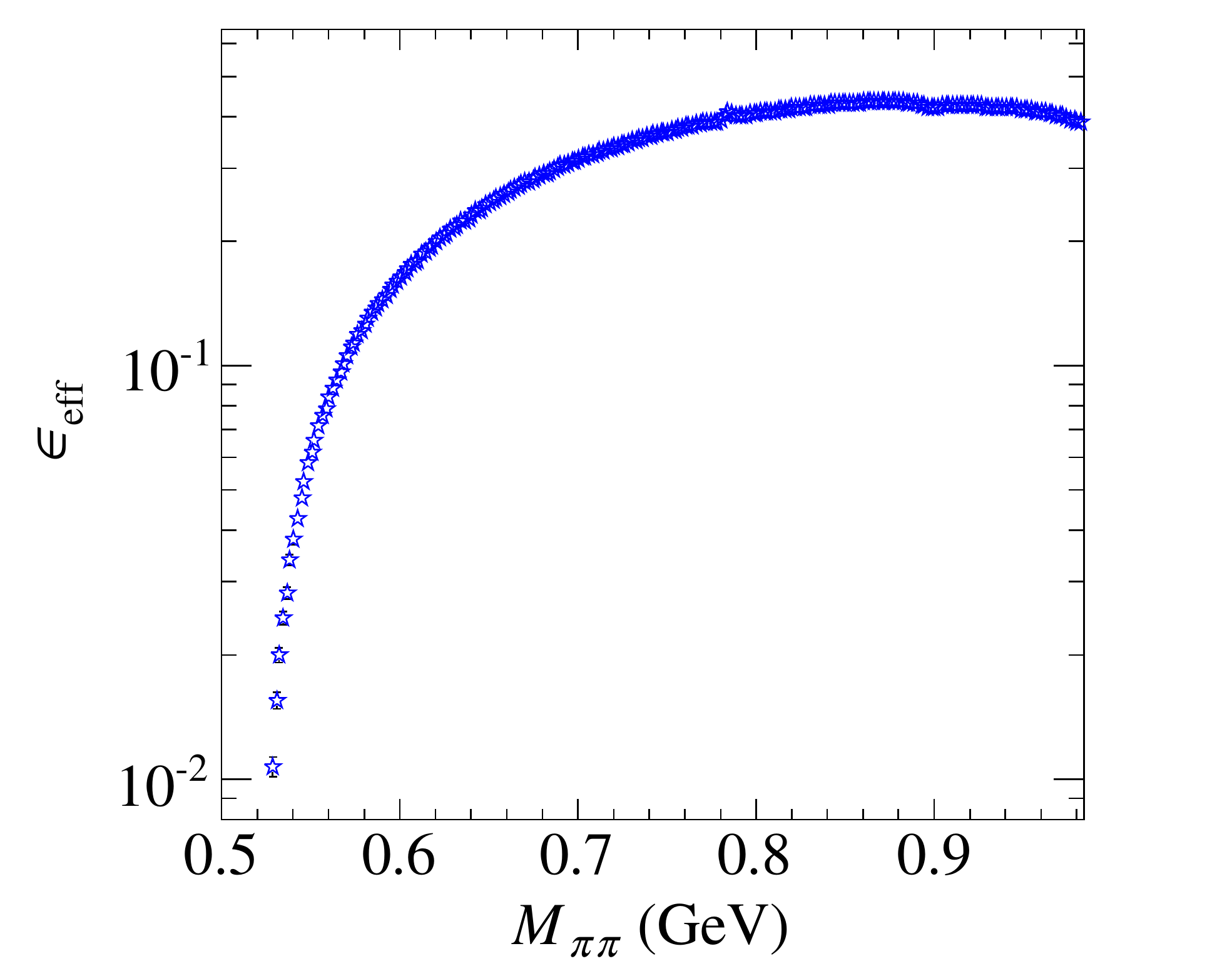}
\caption{Global analysis efficiency as a function of $M_{\pi\pi}$.}
\label{global_eff}
\end{center}
\end{figure}

The total systematic uncertainty due to the background evaluation is shown  in Figure~\ref{tot_tlimit_syst}. 
The full list of the systematic effects taken into account is summarized in Table~\ref{tab:1}. They do not affect the irreducible background estimate performed with the sideband fitting technique, but partially contribute to the background estimate in the $\rho-\omega$ region (see above) and enter in the determination of the selection efficiency and the luminosity measurement.
\begin{table}[h]
\begin{center}
\caption{Summary of the systematic uncertainties affecting the $\piplus \piminus \gamma$ analysis}
\label{tab:1}\begin{small}
\begin{tabular}{cc} 
\hline
Systematic source  & \,\,\,\,\,\,\,\,\,\,\,\,\,\,\,\,\,\,\,\,\,\,\,\,\,\,\,\,\,\,\,\,\, Relative uncertainty (\%)   \\
\hline
\mt\ cut                    & \,\,\,\,\,\,\,\,\,\,\,\,\,\,\,\,\,\,\,\,0.2 \\
Acceptance               & \,\,\,\,\,\,\,\,\,\,\,\,\,\,\,\,\,\,\,\,0.6--0.1 as $M_{\pi\pi}$ increases \\
Trigger                  &\,\,\,\,\,\,\,\,\,\,\,\,\,\,\,\,\,\,\,\,0.1   \\
Tracking                 & \,\,\,\,\,\,\,\,\,\,\,\,\,\,\,\,\,\,\,\,0.3    \\
Generator                 & \,\,\,\,\,\,\,\,\,\,\,\,\,\,\,\,\,\,\,\,0.5 \\
Luminosity               & \,\,\,\,\,\,\,\,\,\,\,\,\,\,\,\,\,\,\,\,0.3    \\
 PID                     &\,\,\,\,\,\,\,\,\,\,\,\,\,\,\,\,\,\,\,\,negligible    \\
Total                 & \,\,\,\,\,\,\,\,\,\,\,\,\,\,\,\,\,\,\,\,0.9--0.7 as $M_{\pi\pi}$ increases\\
 \hline
\end{tabular}
\end{small}
\end{center}
\end{table}
Finally, in Fig.~\ref{global_eff} we show the global analysis efficiency  as estimated with the full  $ \pi^+  \pi^- \gamma$ simulation ($\mathtt{PHOKHARA}$ generator + GEANFI~\cite{GEANFI}).
This includes contributions from kinematic cuts, trigger, tracking, acceptance and PID-likelihood effects. The total systematic error on the global analysis efficiency ranges between 0.7\% and 0.4\% as $M_{\pi\pi}$ increases.

\section{Upper limits}
\label{sec:upper_limit}
\begin{figure}[htp!]
\begin{center}
\includegraphics[width=8cm]{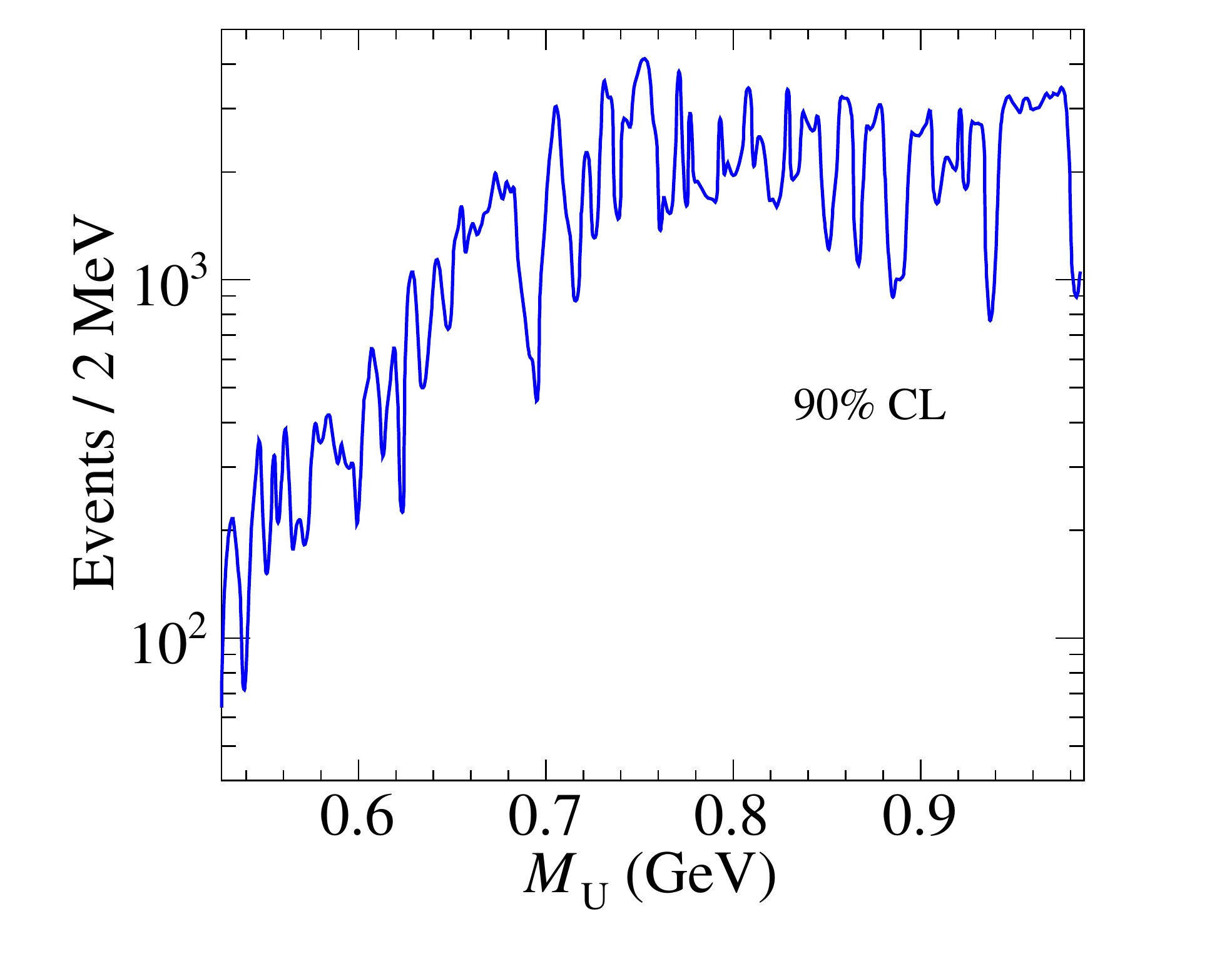}
\caption{Maximum number of U boson events excluded at 90\% CL.}
\label{NCLS}
\end{center}
\end{figure}

We did not observe any excess of events with respect to the estimated background with significance larger than three standard deviations over the whole $M_{\rm U}$ explored spectrum.
We thus extracted the mass dependent limits on $\varepsilon^2$ at 90\% confidence level (CL) by means of the $\mathrm{CL}_\mathrm{S}$ technique~\cite{CLS_Technique}. 
The procedure requires as inputs the measured invariant mass spectrum, the estimated irreducible total background and the U boson signal for each $M_{\pi\pi}$ bin.
The measured spectrum is used as input without any efficiency or background correction.
The signal is generated varying the U boson mass hypothesis in steps of 2~MeV. At each step, a Gaussian peak is built with a width corresponding to the invariant mass resolution of the dipion system.
The systematic uncertainties were taken into account by performing a Gaussian smearing of the evaluated background according to the estimates in Section~\ref{sec:syst}  and Fig.~\ref{tot_tlimit_syst}.
The results of the  statistical procedure are shown in Figure~\ref{NCLS} in terms of the number $N_{\rm CLS}$ of U boson signal events excluded at 90\% CL. 

We computed the limit on the mixing strength $\varepsilon^2$ by means of the following formula~\cite{mmg,eeg}:
\begin{equation}
\varepsilon^2=\frac{\alpha^{\prime}}{\alpha}= \frac{N_{ \rm CLS}/(\epsilon_{\rm eff} \cdot L)}{H \cdot I}.
\label{eq.1}
\end{equation} 
where  $\epsilon_{\rm eff}$ is the global analysis efficiency (see Fig. \ref{global_eff}), $L$ is the integrated luminosity, $H$ is the radiator function computed at QED NLO corrections with a 0.5\% uncertainty~\cite{H,H_1,H_2,H_3}, $I$ is the effective $\mathrm{e}^+\mathrm{e}^- \to \rm U \to \pi^+ \pi^-$ cross section integrated over the single mass bin centered at $M_{\pi\pi}=M_{\rm U}$ with $\varepsilon=1$.
The uncertainties on $H$,  $\epsilon_{\rm eff}$, $L$, and I, propagate to the systematic error on  $\varepsilon^{2}$ via eq.~(\ref{eq.1}). The resulting uncertainty on $\varepsilon^{2}$ is lower than 1\% and has been taken into account in the estimated limit. 
\begin{figure}[h!]
\begin{center}
\hspace{-1.2cm}
\includegraphics[width=9cm]{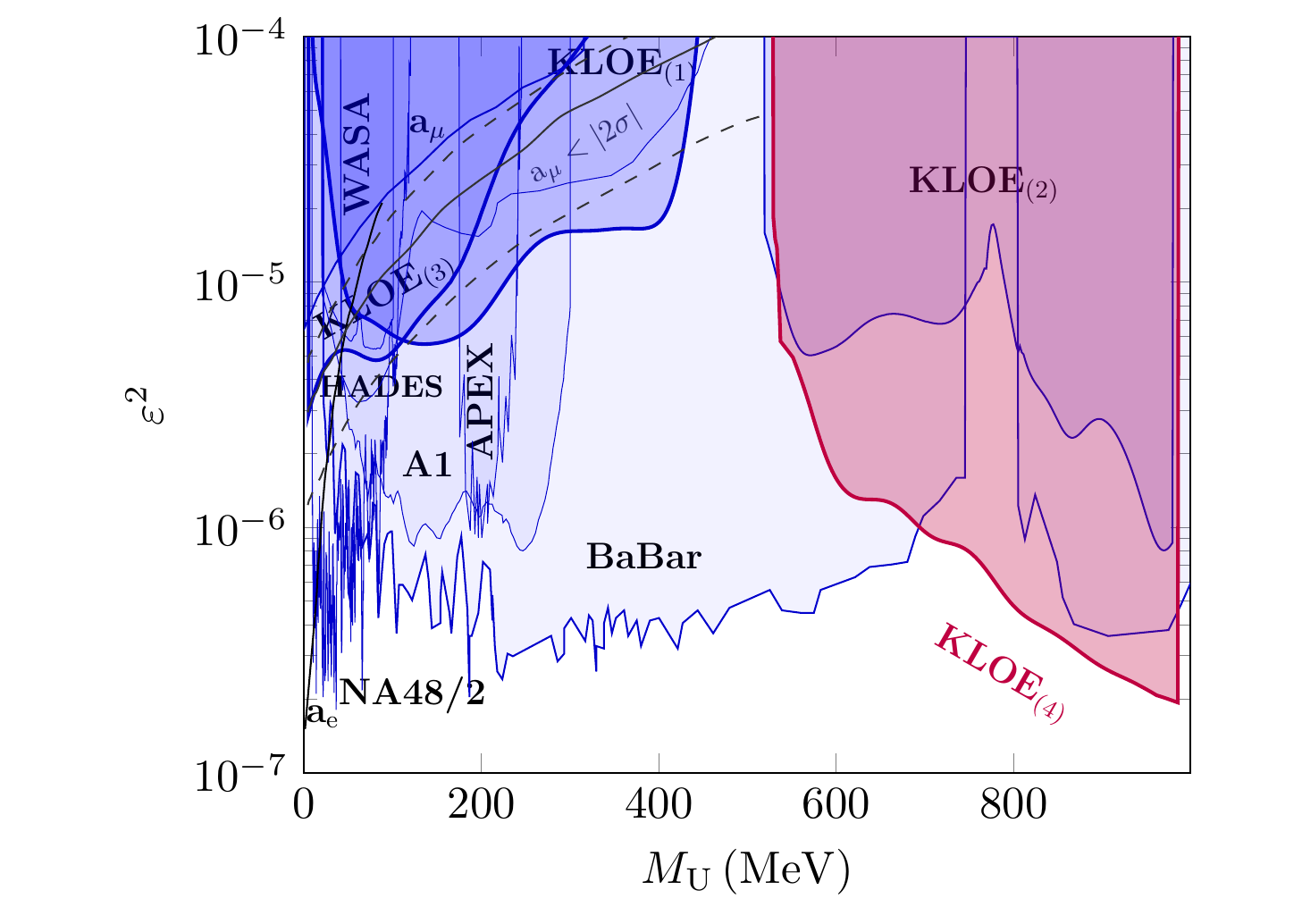}\vspace{-0.5cm}
\caption{
90\% CL exclusion plot  for $\varepsilon^2$ as a function of the $\mathrm{U}$-boson mass (KLOE$_{(4)}$). The limits from  the A1~\cite{Mami1} and APEX~\cite{Apex} fixed-target experiments; the limits from the $\phi$ Dalitz dacay (KLOE$_{(1)}$)~\cite{KLOE_UL1,KLOE_UL2} and $\mathrm{e}^+\mathrm{e}^-\to \mathrm{U}\gamma$ process 
where the U boson decays in $\mathrm{e}^+\mathrm{e}^-$ or  $\mu^+\mu^-$ (KLOE$_{(3)}$ and KLOE$_{(2)}$ respectively)~\cite{eeg,mmg}; the WASA~\cite{WASA}, HADES~\cite{HADES}, BaBar~\cite{BaBar} and NA48/2~\cite{NA48/2} limits are also shown. The solid lines are the limits from the muon and electron anomaly~\cite{a_mu}, respectively. The gray line shows the U boson parameters that could explain the observed $a_{\mu}$ discrepancy with a $2\, \sigma$ error band (gray dashed lines)~\cite{a_mu}.
}
\label{UL}
\end{center}
\end{figure}
Figure~\ref{UL} shows the results from eq.~\ref{eq.1}  after a smoothing procedure (to make them more readable), compared with  limits from other experiments in the mass range 0--1~GeV.
Our 90\% CL upper limit on $\varepsilon^2$ reaches a maximum value of $1.82 \times 10^{-5}$ at 529~MeV and a minimum value of $1.93 \times 10^{-7}$ at 985~MeV. The sensitivity reduction due to the $\omega \to \piplus \piminus \pizero$ decay is of the same order of the statistical fluctuations and  thus not visible after the smoothing procedure.


\section{Conclusions}
\label{sec:conclusions}

We used an integrated luminosity of 1.93 fb$^{-1}$ of KLOE data
 to search for dark photon hadronic decays in the $\eeUgUpp$ continuum process. No signal has been observed and a limit at 90\% CL has been set on the coupling factor $\varepsilon^2$ in the energy range between 527 and 987~MeV. The limit is more stringent than other limits in the $\rhomeson-\wmeson$ region and above.

\section*{Acknowledgments}
\label{sec:acknowledgments}

We warmly thank our former KLOE colleagues for the access to the data collected during the KLOE data taking campaign.
We thank the DA$\Phi$NE team for their efforts in maintaining low background running conditions and their collaboration during all data taking. We want to thank our technical staff: 
G.F. Fortugno and F. Sborzacchi for their dedication in ensuring efficient operation of the KLOE computing facilities; 
M. Anelli for his continuous attention to the gas system and detector safety; 
A. Balla, M. Gatta, G. Corradi and G. Papalino for electronics maintenance; 
M. Santoni, G. Paoluzzi and R. Rosellini for general detector support; 
C. Piscitelli for his help during major maintenance periods. 
This work was supported in part by the EU Integrated Infrastructure Initiative Hadron Physics Project under contract number RII3-CT- 2004-506078; by the European Commission under the 7th Framework Programme through the `Research Infrastructures' action of the `Capacities' Programme, Call: FP7-INFRASTRUCTURES-2008-1, Grant Agreement No. 227431; by the Polish National Science Centre through the Grants No.\
2011/03/N/ST2/02652,
2013/08/M/ST2/00323,
2013/11/B/ST2/04245,
2014/14/E/ST2 /00262,
2014/12/S/ST2/00459.





\end{document}
\endinput